\documentclass[structabstract]{aa}  
\usepackage{graphicx}
\usepackage{natbib}
\usepackage{subfig}
\usepackage{epsfig} 
\usepackage{ifpdf} 
\usepackage{fancyvrb}

\usepackage{txfonts}
\usepackage{xcolor}

\begin{document}
   \title{Origin of the X-ray off-states in Vela X-1}


   \author{A. Manousakis 
          \inst{1,2,3}
          \and
          R. Walter \inst{2,3}
          }
          
               \institute{ 
               	 Centrum Astronomiczne im. M. Kopernika, Bartycka 18, PL-00716 Warszawa, Poland \\
	\email{antonism@camk.edu.pl} 
\and
   ISDC Data Center for Astrophysics, Universit\'e de Gen\`eve, Chemin d'Ecogia 16, CH-1290 Versoix, Switzerland  
        \and Observatoire de Gen\`eve, Universit\'e de Gen\`eve,  Chemin des Maillettes 51, CH-1290 Versoix, Switzerland        
             }

   \date{Received Feb. XX, 2013; accepted Dec XX, 2013}

 
  \abstract
   {Vela X-1 is the prototype of the classical sgHMXB systems. Recent continuous and long monitoring campaigns revealed a large hard X-rays variability amplitude with strong flares and off-states. This activity has been interpreted invoking clumpy stellar winds and/or magnetic gating mechanisms.}
   { We are probing if the observed behaviour could be explained by unstable hydrodynamic flows close to the neutron star rather than the more exotic phenomena.}     
   {We have used the hydrodynamic code VH-1 to simulate the flow of the stellar wind with high temporal resolution and to compare the predicted accretion rate with the observed light-curves.}
   {The simulation results are similar to the observed variability. Off-states are predicted with a duration of 5 to 120 minutes corresponding to transient low density bubbles forming around the neutron star. Oscillations of the accretion rate with a typical period of $\sim$ 6800 sec are generated in our simulations and observed. They correspond to the complex motion of a bow shock, moving either towards or away from the neutron star. Flares are also produced by the simulations up to a level of $10^{37}$ erg/s.}
   {We have qualitatively reproduced the hard X-ray variations observed in Vela X-1 with hydrodynamic instabilities predicted by a simple model. More sophisticated phenomena, such as clumpy winds or the magnetic gating mechanism, are not excluded but not required to explain the basic phenomenology.}%



\maketitle
%

\section{Introduction}
\label{sec:intro}

Neutron stars or black-holes in high-mass X-ray binaries (HMXBs) accrete gas from the stellar wind of their massive, OB-type, stellar companions. A fraction of the gravitational potential energy is converted into X-rays, ionizing and heating the surrounding gas. The X-ray emission can be used to investigate the structure of the stellar wind \emph{in situ} \citep{Walter07Winds}.

Vela X-1 (=4U 0900$-$40) is a classical persistent and eclipsing super-giant High Mass X-ray Binary (sgHMXB). 
The system consists of an evolved B 0.5 Ib supergiant (HD 77581) and of a massive neutron star 
\citep[M$_{NS}=1.86$ M$_{\odot}$;][]{Quaintrell_et_al03}. 
The neutron star orbits its massive companion with a period of about 8.9 days,
 in a  circular orbit \citep[$e\approx 0.09$;][]{1997ApJS..113..367B} with a radius of $\alpha=$1.76 $R_{*}$. The  stellar wind is characterized by a mass-loss rate of $\sim 4\times10^{-6}$ M$_{\odot}$ yr$^{-1}$ \citep{1986PASJ...38..547N} and a wind terminal velocity of $\upsilon_{\infty}\approx 1700$ km s$^{-1}$ \citep{1980ApJ...238..969D}. The X-ray luminosity is typically $\sim 4 \times 10^{36}$ erg s$^{-1}$, although high variability can be observed \citep{1999A&A...341..141K}.

Recent studies on the hard X-ray variability of Vela X-1 have revealed a rich phenomenology including flares and short off-states \citep{Kreykenbohm+08}.
 Both flaring activity and off-states were interpreted as the effect of a strongly structured wind. \citet{Furstetal10} characterized the X-ray variability of Vela X-1 with a log-normal distribution, interpreted in the context of a clumpy stellar wind.  Off-states have been interpreted \citep{Kreykenbohm+08}  as an evidence for the propeller effect \citep{1975A&A....39..185I}, possibly accompanied by leakage through the magnetosphere \citep{2011A&A...529A..52D}. The quasi-spherical subsonic accretion model \citep{2012MNRAS.420..216S,2013MNRAS.428..670S} is an alternative, predicting that the repeatedly observed ‘off-states’ in Vela X-1 are the result of a transition from  Compton to radiative cooling (higher and lower luminosity, respectively).

In this paper we present new results from 2-D hydrodynamic simulations of Vela X-1 and conclude 
that the observed phenomenology can be explained qualitatively without intrinsic clumping or the propeller effect. The code and the simulations are described in  Sect. \ref{sec:hydro}. The simulation results are presented and compared to observations in  Sect. \ref{sec:results}, and discussed in Sect. \ref{sec:discussion} and \ref{sec:conclude}.


\section{Hydrodynamic Simulations} 
\label{sec:hydro}

 \subsection{The hydrodynamic code}

The motion of the fluid is described by the Euler equations assuming mass, momentum, and energy conservation. The internal energy in each cell is described by the first law of thermodynamics. The following set of equations are therefore solved in a fixed, non-uniform mesh:

\begin{equation}
\partial_{t}\rho +\nabla \cdot (\rho \mathbf{u})=0
\end{equation}

\begin{equation}
\partial_{t}(\rho \mathbf{u} )+ \nabla \cdot (\rho \mathbf{u} \mathbf{u}  )
  +\nabla P = \mathbf{F} 
\end{equation}

\begin{equation}
\partial_{t}  E +\nabla \cdot (
 E \mathbf{u}  +  P \mathbf{u} ) =  \mathbf{u} \cdot \mathbf{F}
\end{equation}
where we used $\partial_{t}=\partial/ \partial t$. 
The primary variables of the simulations (the mass density $\rho$, gas pressure $P$, and fluid velocity $\mathbf{u}$) are  
completed by the total energy  $E=\frac{1}{2} \rho \mathbf{u}^{2}+\rho e$ and the force $\mathbf{F}$, accounting for the Roche potential and the line driven force (see Sect. \ref{sec:sw}). The specific internal energy ($e$) is related to the pressure through the equation of state, $P=\rho e (\gamma - 1)$, where $\gamma=5/3$ is the ratio of specific heats.

The  VH-1\footnote{http://wonka.physics.ncsu.edu/pub/VH-1/} hydrodynamic code is described in detail in \citet{Blondin90,Blondin91}. The simulations of Vela X-1 take into account the gravity of the primary and of the neutron star, the radiative acceleration \citep[CAK hereafter]{CAKwind} of the stellar wind of the donor star, and the suppression of the stellar wind acceleration due to high ionization within the Str\"{o}mgren sphere of the neutron star. The parameters are listed in table \ref{tab:VELAparams}.

The equations are solved in the orbital plane of the co-rotating reference system, where the  lateral extent ($\theta$ component) of the spherical mesh is 
only one cell. 
We have also assumed a circular orbit and a synchronous rotation. The code uses the piecewise parabolic method for shock hydrodynamics developed by \citet{PPMCW}.

A computational mesh of 900 radial by 347 angular zones, extending  from 1 to $\sim$ 25 R$_{*}$ and in 
angle from $-\pi$ to $+\pi$, has been employed. The grid is built in a non-uniform way in order to allow for higher resolution (up to $\sim 10^{9}$ cm) towards the neutron star and centred on the center of mass.

The initial wind density, velocity, pressure and CAK parameters of the 2D simulartion are set by the results of a 1D CAK/Sobolev simulation (also extending up to 25 R$_{*}$) of a single star and resulting in the standard $\beta\approx 0.8$ velocity law. In the binary potential, the wind takes a few days of simulation to relax and find a new equilibrium. The first 3 days of the simulations are therefore excluded from the variability analysis. 

The wind reaching the radial outermost part of the mesh is characterized as an outflow boundary condition i.e. it leaves the simulation domain. We also assume that the wind entering the cell representing the surface of the neutron star is completely accreted, leaving an (almost) zero density and pressure. The matter falling from the surrounding cells is therefore in free fall \citep{1971MNRAS.154..141H,2009ApJ...700...95B}.

The time step of the simulations is $\sim 1/10$ sec and the variables of each cell are stored for each step. The code also calculate the ionization parameter \citep[$\xi= L_{X}/n r_{ns}^2$, where $n$ is the number density at the distance $r_{ns}$ from the neutron star;][]{1969ApJ...156..943T} and the instantaneous mass and angular momentum accreted ($\dot{M}_{acc}$) on the neutron star. The code was run for about 30 days, i.e. more than three orbits. This is enough for the wind to reach a stable configuration and to study its short time scale variability.

\subsection{Stellar wind acceleration}
\label{sec:sw}
The winds of hot massive stars are characterized observationally by the wind terminal velocity and the mass-loss rate. The velocity  is  described by the  $\beta$-velocity law, $\upsilon=\upsilon_{\infty}(1-R_{*}/r)^{\,\beta}$, where $\upsilon_{\infty}$ is the terminal velocity and $\beta$ is the gradient of the velocity field. 
For supergiant stars, values for wind terminal velocities and mass-loss rates are in the range $\upsilon_{\infty}\sim 1500-3000$ km s$^{-1}$ and $\dot{M}_{\rm w}\sim 10^{-(6-7)}$ M$_{\odot}$ yr$^{-1}$, respectively \citep{winds_from_hot_stars}.   
 
The  stellar winds  of  massive supergiant stars are  radiatively driven by absorbing ultraviolet photons from the underlying photosphere. To properly simulate the radiation force driving the stellar wind, we used the CAK/Sobolev approximation:
\begin{equation}
\label{eq:radforce}
F_{\rm rad}=\frac{\sigma_{e}L_{*}}{4\pi c R^{2}} k K_{FDC} \Big(
\frac{1}{\sigma_{e} \rho u_{th}} \frac{du}{dR}
\Big)^{\alpha},
\end{equation}
where $L_{*}$ is the stellar luminosity, $\sigma_{e}$ is the electron scattering 
coefficient ($\approx 0.33$ cm$^{2}$g$^{-1}$) and $u_{th}$ is the thermal velocity of the gas. The parameters CAK-$k$ and CAK-$\alpha$ are constants and correspond to the number and strength of the absorption lines, respectively \citep{CAKwind}.
The effect of finite disk correction \citep{1986ApJ...311..701F}  has been accounted through the factor
\begin{equation}
K_{\rm FDC}=\frac{(1+\sigma)^{1+\alpha}-(1+\sigma\mu^{2})^{1+\alpha}}
{(1+\alpha)(1-\mu^{2})\sigma(1+\sigma)^{\alpha}},
\end{equation}
where $\sigma=\frac{dln\, u}{dln\, R}-1$ and $\mu=\big(1-\frac{R_{*}^{2}}{R^{2}}\big)^{1/2}$, 
 where R and u are the radial distance and velocity, respectively.
The finite disk correction produces a shallow $\beta\approx0.8$ velocity law, corresponding to the observations, rather than a steeper $\beta\approx 0.5$. The wind parameters, including the density at the bottom of the donor star atmosphere ($\rho_{0}$), and the resulting mass-loss rate and wind terminal velocity are listed in table \ref{tab:VELAparams}.

The radiation force is known to be unstable \citep{1984ApJ...284..337O}, generating inhomogeneities and clumps \citep{Owocki+88,2013MNRAS.428.1837S}. 
In a close binary system the neutron star is a driver of the hydrodynamics and models combining binarity and intrinsic instabilities are still to be developed.
  
\begin{table}[h] 
\caption{Parameters of the simulations.}
\label{tab:VELAparams}

\centering                          
\begin{tabular}{p{6cm} l c }        
\hline
\hline                 
Parameter 	&    Value\rule{0pt}{2.6ex}               \\    
\hline        
\hline
\emph{Donor star Parameters\rule{0pt}{2.6ex}} & \\
M$_{*}$ & 23.1 M$_{\odot}$  \\
R$_{*}$ & 30 R$_{\odot}$   \\
L$_{*}$ & $2.5\times 10^{5}$ L$_{\odot}$\\  
T$_{*}$ & 40000 K  \\
\hline
\emph{Binary parameters\rule{0pt}{2.6ex}} & \\
M$_{NS}$  & 1.86 M$_{\odot}$    \\
$\alpha$  & 1.76 R$_{*}$      \\
L$_{X}$   & $4\times 10^{36}$ erg s$^{-1}$  \\
\hline                                   

\hline
\emph{CAK parameters\rule{0pt}{2.6ex}} & \\
CAK-$\alpha$  & 0.58 \\
CAK-$k$       &  0.80 \\
$\rho_{0}$    &  $10^{-11}$ g cm$^{-3}$\\

\hline
\emph{Wind Parameters\rule{0pt}{2.6ex}} & \\
$\dot{\rm{M}}_{W}$& $4\times 10^{-6}\, {\rm M}_{\odot}$ yr$^{-1}$   \\
$\upsilon_{\infty}$ & 1700 km s$^{-1}$   \\
\hline                                   
\end{tabular} \newline

\end{table}

The radiative acceleration of the wind is suppressed in case of X-ray photo-ionization. The effects of the X-ray radiation on the radiative acceleration force are complicated due to the large number of ions and line transitions contributing to the opacity \citep{1982ApJ...259..282A,1990ApJ...365..321S}. Detailed NLTE wind models of the envelope of  Vela X-1 shows the existence of a photoionized `bubble' around the neutron star filled with stagnating flow \citep{2012ApJ...757..162K}.

Assuming that the gas is in ionization equilibrium, the ionization state can be estimated with the $\xi$ parameter \citep{FarnssonFabian1980,Blondin90}. In our simulations we have defined a critical ionization parameter $(\xi>10^{2.5}$ erg cm sec$^{-1})$ above which most of the elements responsible for the wind acceleration (e.g., C, N, O) are fully ionized \citep{Kallman82} and the radiative force becomes negligible ($F_{\rm rad}=0$ in  Eq. \ref{eq:radforce}).

The main effect of the ionization is the  reduction of the  wind velocity in the vicinity of the neutron star and therefore the enhancement of the mass accretion rate onto the compact object (further increasing the effect). The outcomes of our simulations are not significantly affected by small variations ($\sim$ 20\%) of the critical ionization parameter. However, larger variations (order of magnitude) significantly changes the hydrodynamics. The suppression of the acceleration also triggers the formation of a dense wake at the rim of the Str\"{o}mgren zone \citep{FarnssonFabian1980}, and  has an impact on  the absorption at late orbital phases. X-ray ionization can also affect the thermal state of the wind through X-ray heating and radiative cooling. Such effects are not included in our simulations. 


\section{Results} 
\label{sec:results}

The observed hard X-ray light-curves of Vela X-1 were obtained form the INTEGRAL \citep{integral-ref} soft $\gamma$-ray imager ISGRI  \citep{isgri-ref} in the 20-60 keV energy band and from the PCA \citep{PCA-ref2} detector on board RXTE in the 10-50 keV energy band (lower energies were excluded not to be affected by the variable absorption). The light-curves were obtained using the HEAVENS\footnote{http://www.isdc.unige.ch/heavens} interface \citep{heavens-ref}.  The temporal resolutions of the ISGRI and PCA lightcurves are $\sim$ 40 mins and 1 min, respectively. We excluded data obtained during eclipses, using the orbital solution derived from \citet{Kreykenbohm+08}. 

\begin{figure}
\centering
\includegraphics[width=0.5\textwidth,angle=0]{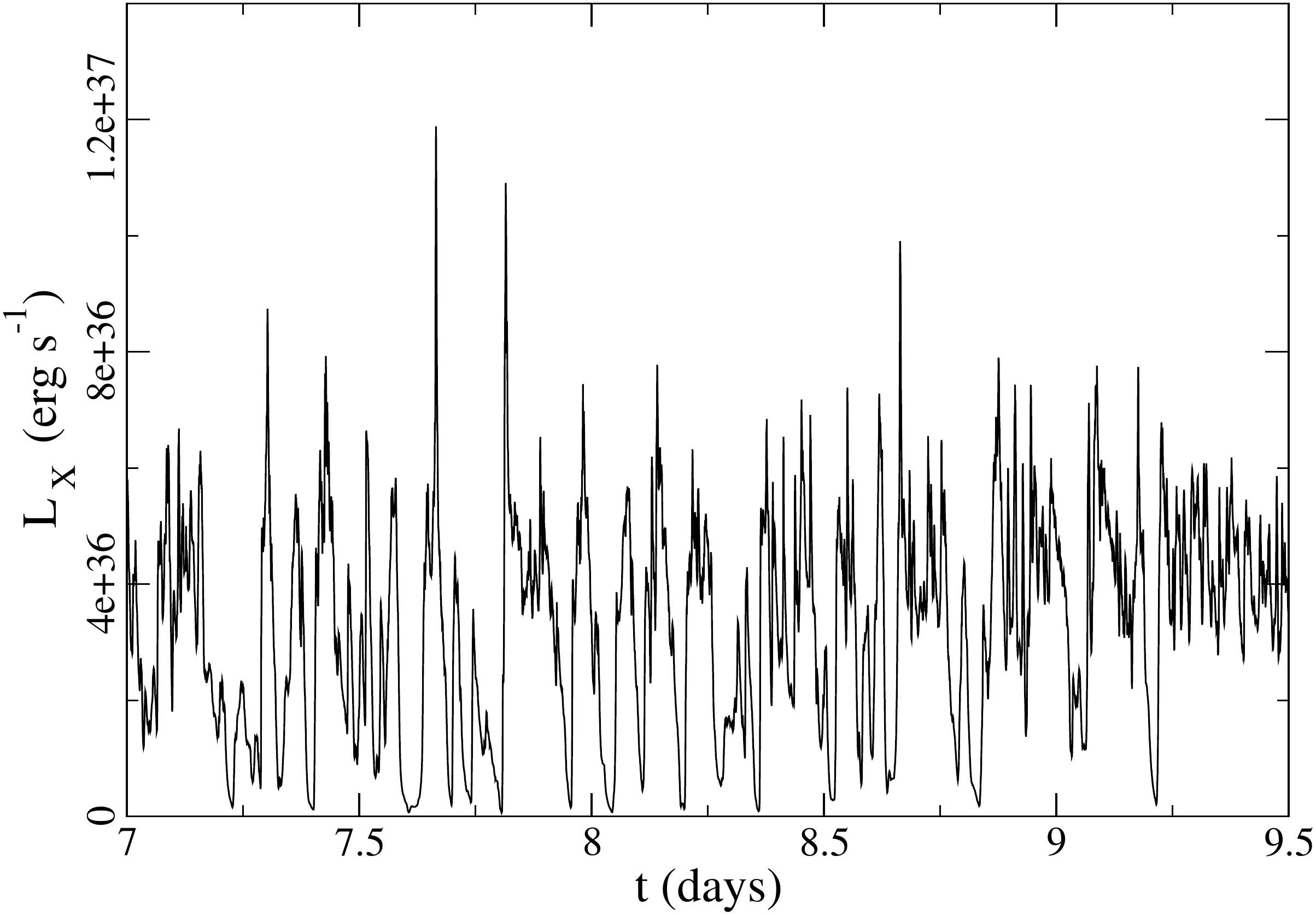} 
\caption{ A fraction of the simulated light-curve, spanning 2.5 days (about 30\% of an orbit).}
\label{fig:velaLC}
\end{figure}

The simulated X-ray lightcurve has been obtained from the instantaneous mass accretion rate (L$_{acc}=\eta \dot{\rm M}_{acc}$ c$^{2}$), using a radiative efficiency of $\eta \approx 0.1$. Figure \ref{fig:velaLC} shows a fraction of the simulated light-curve. A number of off-states can be observed as well as flares reaching $\ga 10^{37}$ erg s$^{-1}$. 
 Figure \ref{fig:offstates} shows a zoom on one of the off-states, where the light-curve has been convolved  with a sinusoidal of  period of 283 sec to account 
for the spin of the Vela X-1 pulsar. 
The properties of the simulated lightcurve and its comparison with the observations are described in the next sub sections.

\begin{figure}[h]
\centering
\includegraphics[width=0.5\textwidth,angle=0]{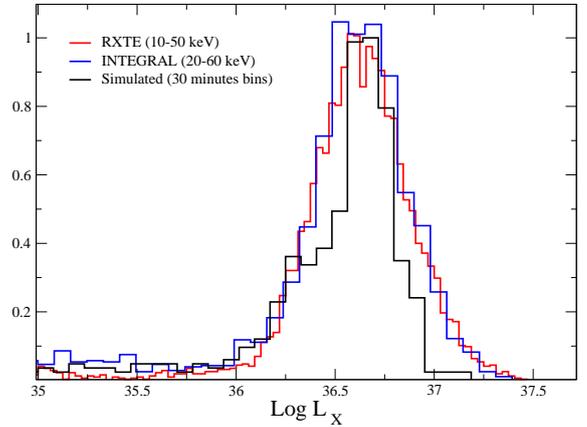} 
\caption{ The X-ray luminosity  distribution of Vela X-1 from RXTE (10-50 keV; red) and INTEGRAL (20-60 keV; blue) observations together with that derived from the simulations (black). The simulated lightcurve was re-binned to 30 minutes bin size to match the INTEGRAL data.}
\label{fig:lognorm}
\end{figure}

\subsection{Luminosity distributions}

We have constructed histograms of the observed and simulated luminosities using the complete  sample available (excluding the first 3 days of the simulations and the eclipses from the observations).  The observed luminosity were derived from the X-ray count rates and responses assuming a distance of $1.9$ kpc \citep{1989PASJ...41....1N}, 
while $\eta$ was adjusted to the value of 0.107 in order to match the averaged luminosity of the simulation to the observed ones. The histograms are characterized by a peak at $\sim 4\times10^{36}$ erg s$^{-1}$ (Fig. \ref{fig:lognorm}) and normalized to the same maximum amplitude for comparison.

The two histograms derived from the observations are very similar and shaped as a log-normal distribution with a low luminosity tail. The log-normal standard deviation is $\sigma\approx0.23$. The histogram of the 30 minutes binned simulated light-curve is characterized by a narrower distribution $(\sigma=0.18)$, however more realistic simulations (e.g., including 3D, heating, cooling and clumping effects)   may generate more turbulence and match better the observations.
Note that the histogram of the un-binned simulated lightcurve shows increased excess (by 25\%) at low luminosity (Log L$_{X}\sim 35.5$).

\subsection{Off states}
\label{sec:offstate}
The simulated light-curves feature a large number of off-states. We defined an off-state when the instantaneous X-ray luminosity dropped below 1/10 of the average luminosity, i.e. $L_{off}\la 4\cdot 10^{35}$ erg s$^{-1} \approx 0.1 \langle L_{X}\rangle$, which is approximately the sensitivity limit of ISGRI in 20 sec.

\begin{figure*}
\centering
\hspace{-1cm}
\includegraphics[width=0.5\textwidth,angle=0]{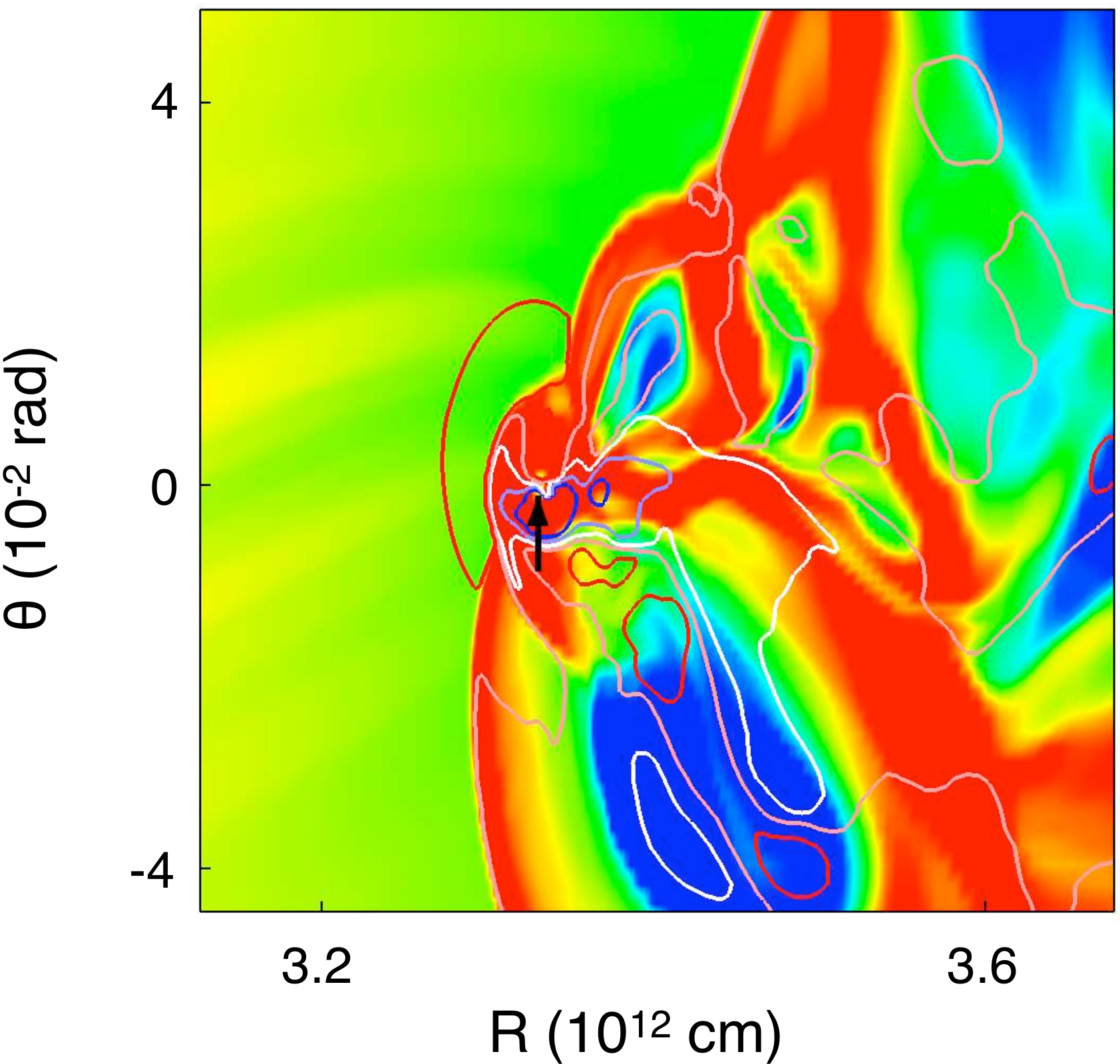} 
\includegraphics[width=0.5\textwidth,angle=0]{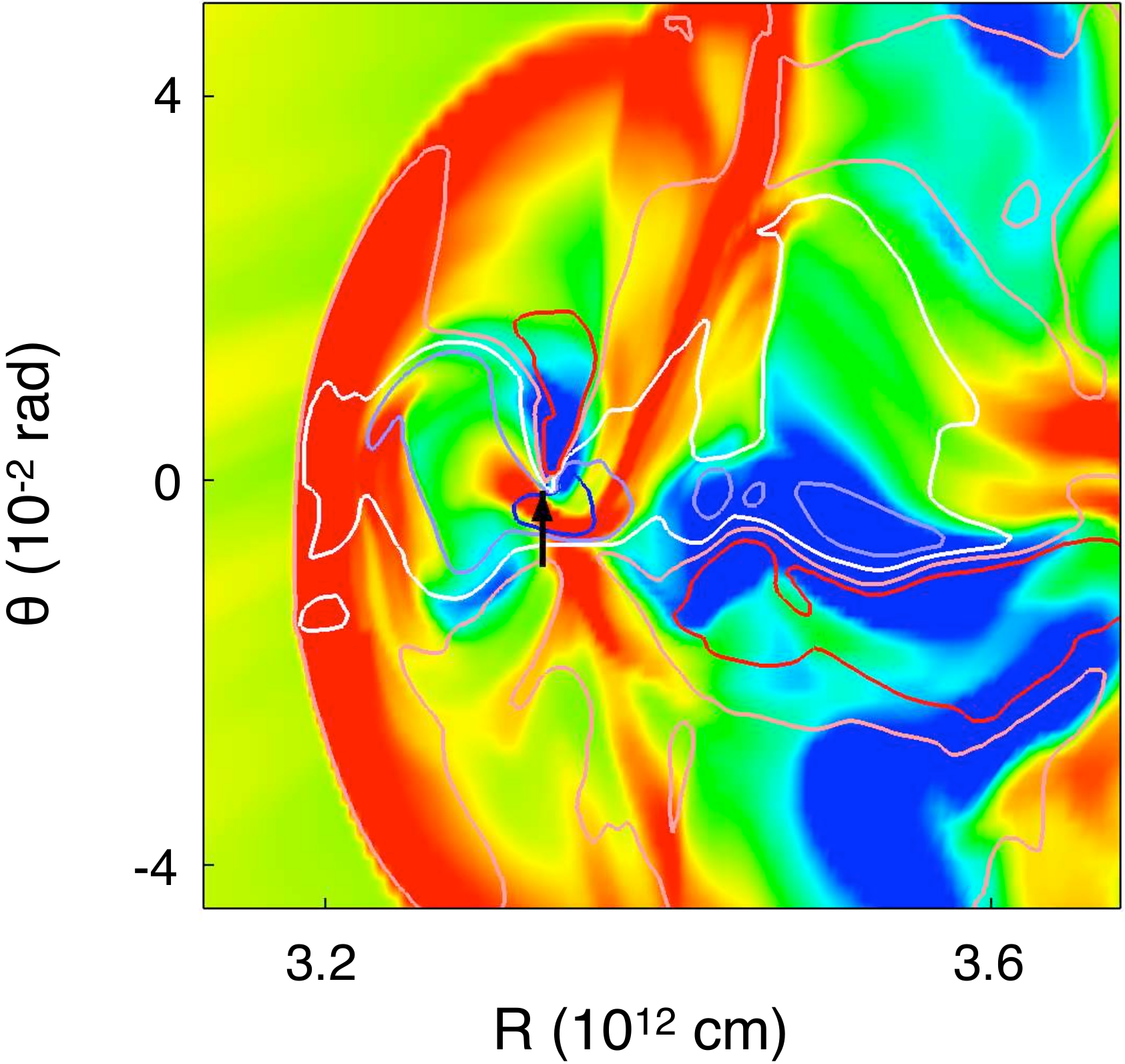} 
\\
\vspace{+0.15cm} \hspace{-0.9cm}
\includegraphics[width=0.5\textwidth,angle=0]{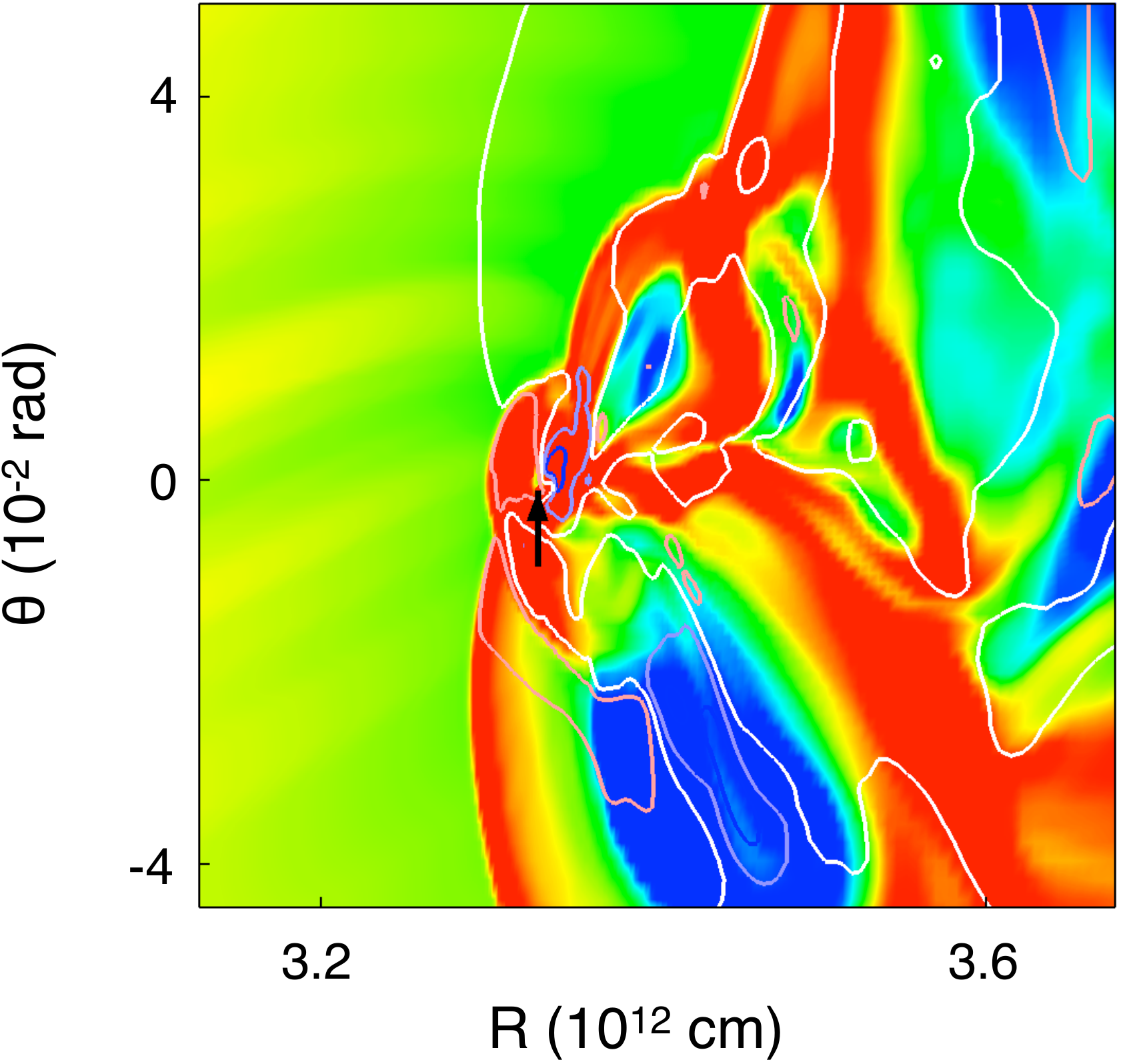}  
\includegraphics[width=0.5\textwidth,angle=0]{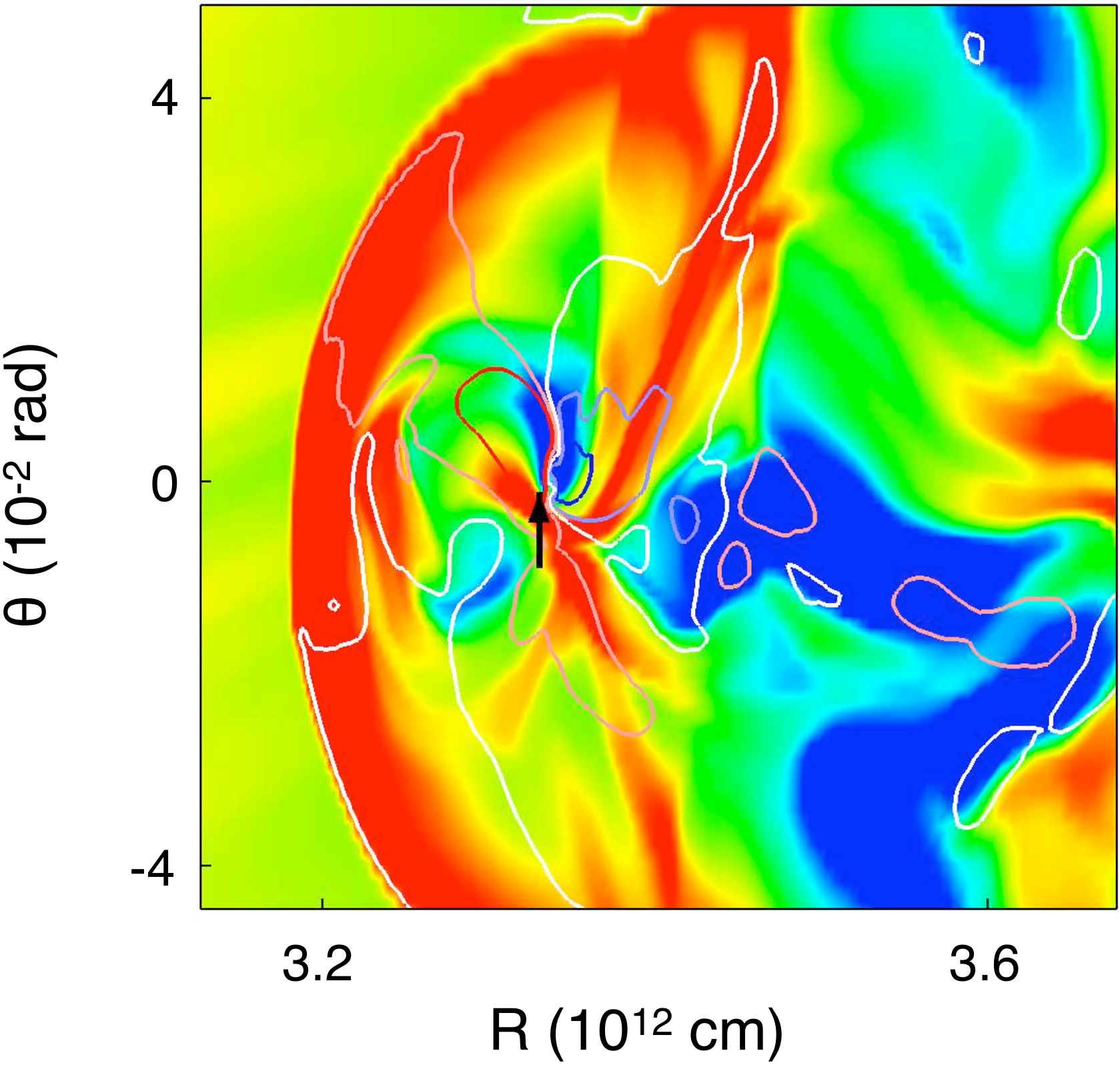} 
\\
\vspace{+0.15cm}
\includegraphics[width=0.175\textwidth,angle=0]{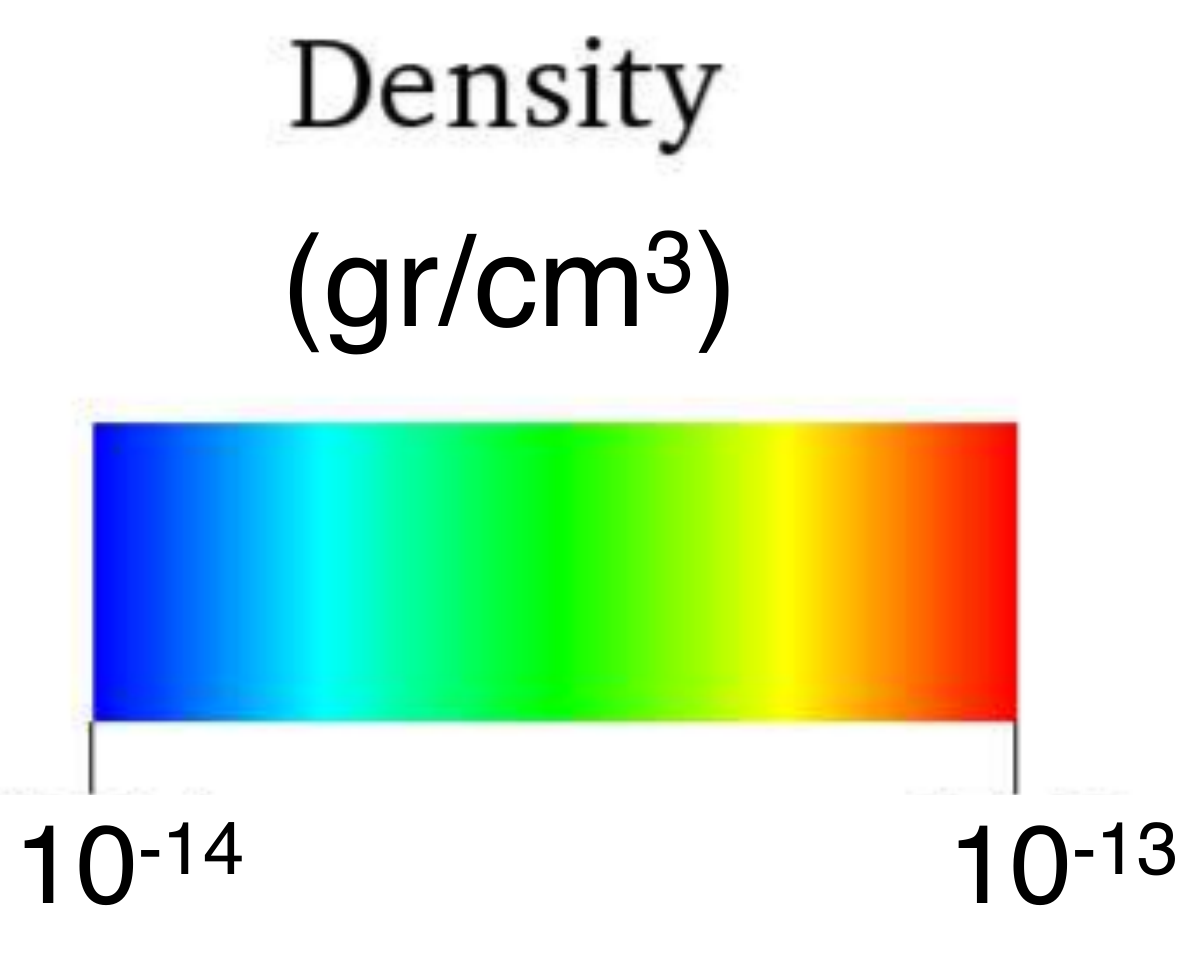} \hspace{2cm}
\includegraphics[width=0.155\textwidth,angle=0]{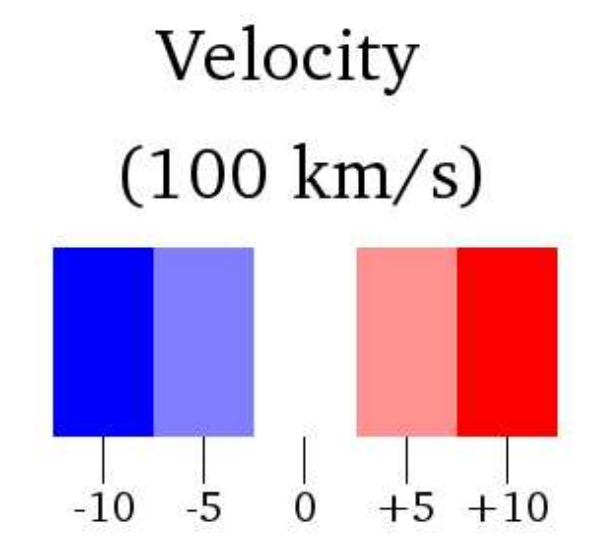} 

\caption{Density distribution (in gr cm$^{-3}$) before (left columns) and during (right columns) the off-state. The upper and lower panels shows the radial and angular velocity contours, respectively. The position of the neutron star is indicated by the black arrow.}
\label{fig:map_offstate}
\end{figure*}

Figure \ref{fig:map_offstate} shows the density maps and the velocity contours before and during an off-state. The velocity contours are shown for the radial and angular velocities (upper and lower panels, respectively). As the bow shock on the left of the neutron star expands, the mass-accretion rate decreases and an off-state occurs. 

The typical size of the bubble sustained by the shock is of the order of $10^{11}$ cm. Inside the density drops at least by a factor of $\sim 10$ when compared with the time-averaged density. This reduced density leads to the chop of the X-ray emission. The duration of the off-state varies with the size of the bubble.

\begin{figure}
\centering
\includegraphics[width=0.45\textwidth,angle=0]{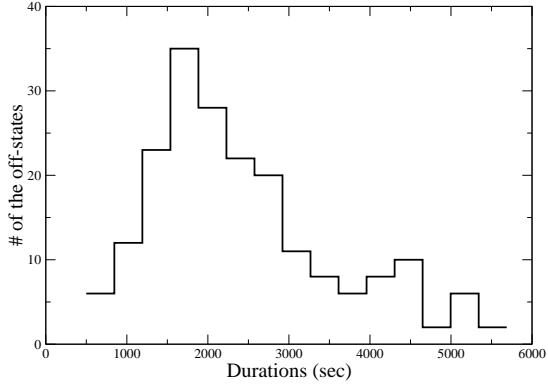} 
\caption{Histogram of the duration of the off-states in the complete simulated lightcurve.}
\label{fig:duroffstate}
\end{figure}

Figure \ref{fig:duroffstate} shows the distribution of the duration of the off-states in the simulated light-curve.  
The typical duration of most of the off-states is about 30 minutes and ranges from 10 minutes to about two hours.  Although the number of observed off-states is small, 
all  of them lasts for between 5 and 30 minutes which coincides well with the simulations.

\begin{figure}
\includegraphics[width=0.45\textwidth]{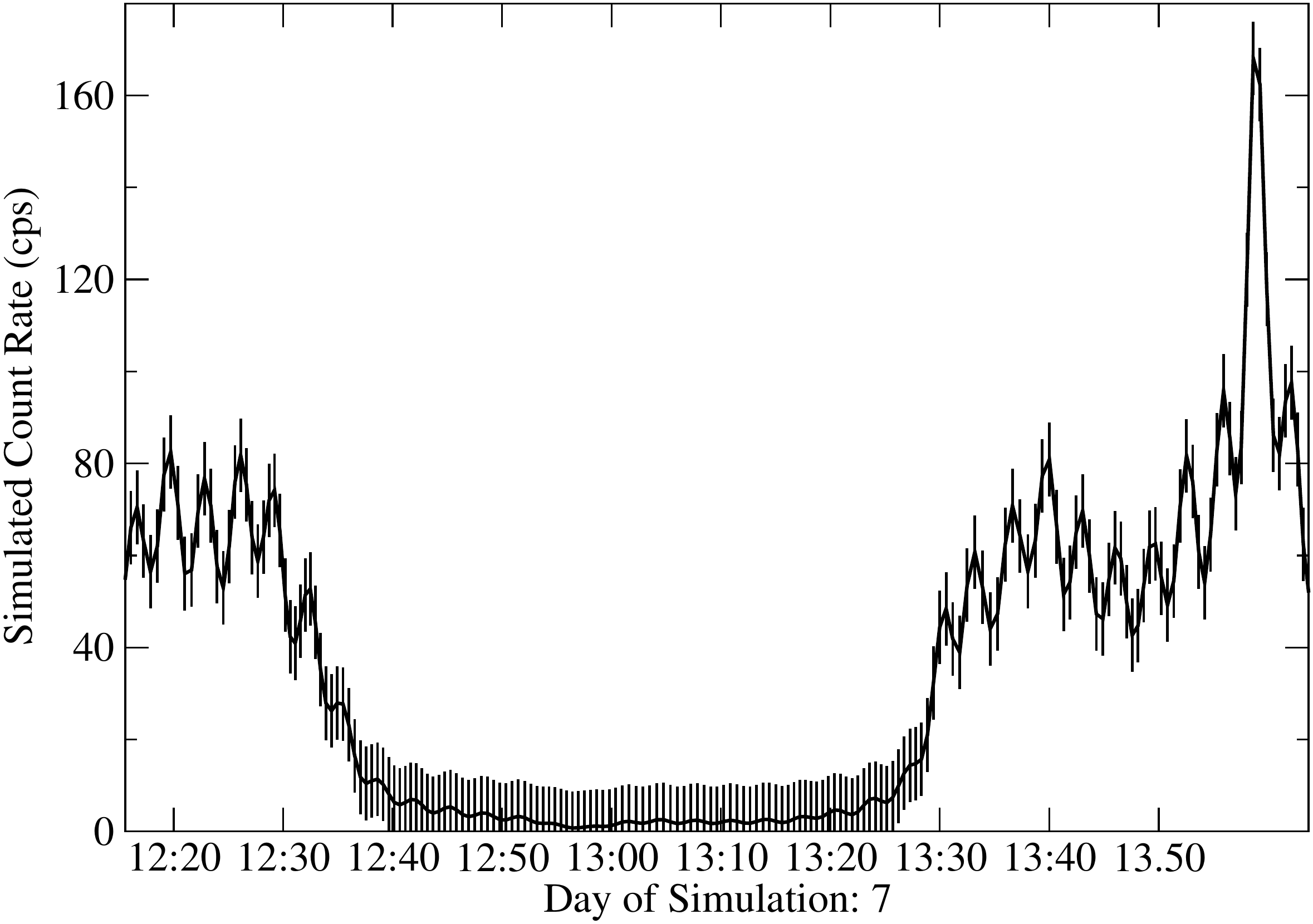} \\
\caption{ A portion of the simulated light-curve with 20 sec time bins.    
The average luminosity of  $\sim 4 \times 10^{36}$ erg s$^{-1}$ corresponds to $\sim 70$ cps. This figure 
can be directly compared to Fig. 6 in \citet{Kreykenbohm+08}. }
\label{fig:offstates}
\end{figure}

\subsection{Flaring activity}

In addition to the off-states we identified prominent flares during the simulations. The flares could reach luminosity up to $L_{flare}\ga 10^{37}$ erg s$^{-1}$ for a duration of $\sim 5-30$ minutes. Figure \ref{fig:offstates} shows the brightest flare reaching $\sim 1.2\times 10^{37}$ erg s$^{-1}\approx 3 \langle L_{X} \rangle$ and lasting for $\sim 25$ minutes. The energy released during this flare is $\sim 10^{39}$ ergs. These flares are compatible, in terms of dynamical range with the flares of Vela X-1 observed by INTEGRAL \citep{Kreykenbohm+08}  however they  are usually shorter, explaining why the luminosity distribution is narrower (see Fig. \ref{fig:lognorm}).  We identified 8 such flares (L$_{X}>10^{37}$ erg s$^{-1}$) in the 30 days simulation while \cite{Kreykenbohm+08} detected 5 flares during an observation of about two weeks.

\subsection{Quasi-Periodicity}

\begin{figure}
\centering
\includegraphics[width=0.45\textwidth,angle=0]{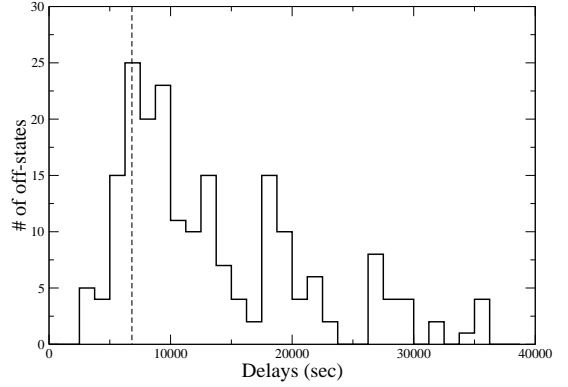} 
\caption{Histogram of delay between two subsequent off-states  using all the available simulated data. A total of 201 off-states have been identified. The dashed vertical line indicates the 6800 sec quasi-periodicity detected in the observations.}
\label{fig:LS6820}
\end{figure}

\begin{figure}
\centering
\includegraphics[width=0.4\textwidth,angle=0]{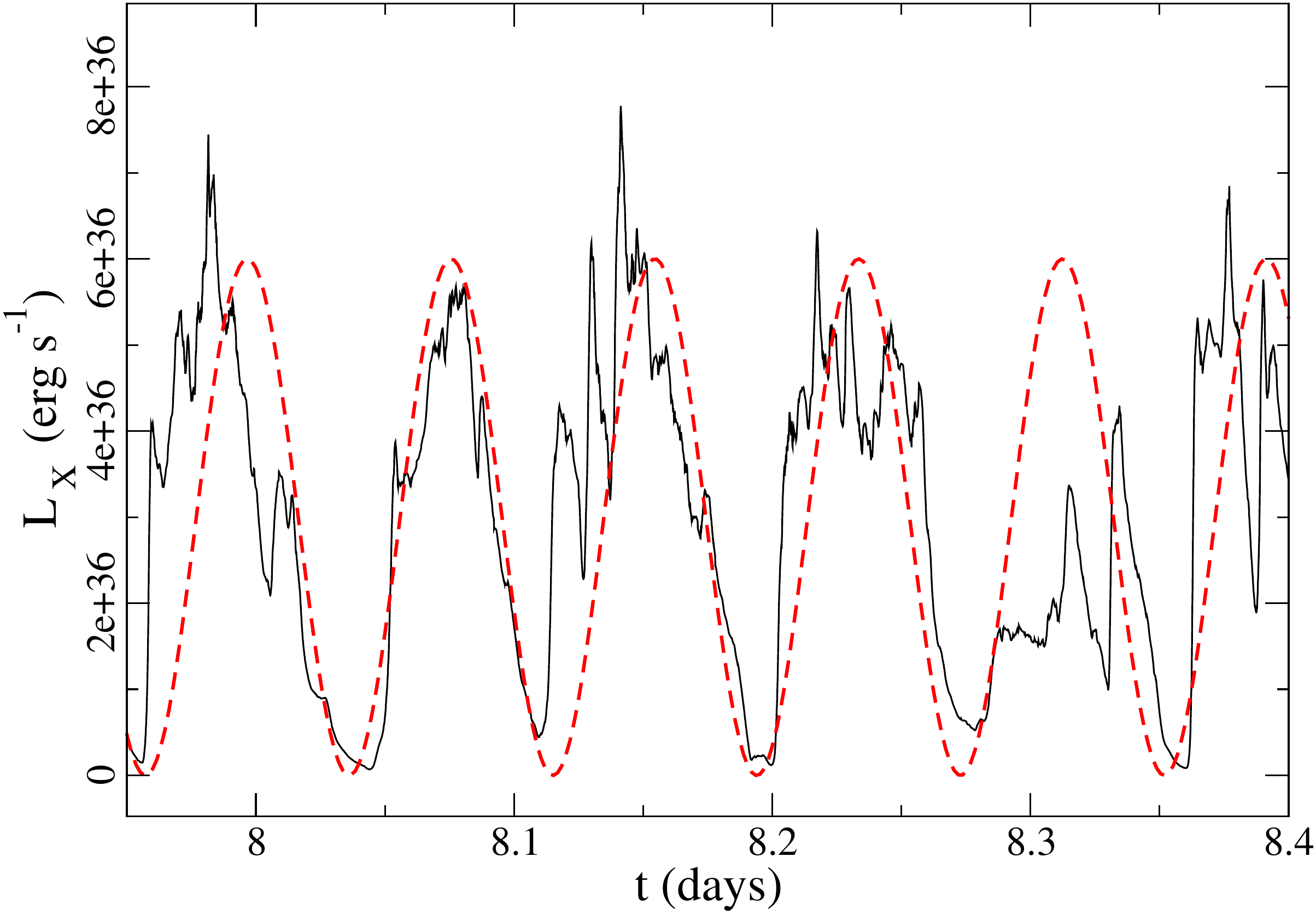} 
\caption{ A section of the simulated light-curves of Vela X-1 together with a 
sine wave of period $6820$ sec (red dashed line).}
\label{fig:vela6820modulo}
\end{figure}

Although the simulated light-curve lacks any periodic signal, we can identify a likely quasi-periodic behaviour related to the spacing of off-states.  Figure \ref{fig:LS6820} shows the histogram of the time intervals between successive off-states. The distribution peaks in the range 6500-7000 sec, close to the transient period of $\sim$ 6800 sec    detected  with INTEGRAL by \citet{Kreykenbohm+08} during $\sim$ 10 hours.
The off-states labelled 1,\, 2,\, 3 and 5 in \citet{Kreykenbohm+08}, which reach less than 10 ct/s, are within $\sim$ 0.1 in phase of the minima of 
the extension of the modulation mentioned earlier. It is therefore plausible that the observed off-states and modulations are the signature of a single physical mechanism 
driving the variability. In our simulations, these modulations last typically for $8 - 16$ hours and repeat every few days.
Some signal is also detected at multiples of that period suggesting that the density of the bubbles does not always reach the threshold we have  defined for an off-state.
A section of the simulated lightcurve featuring a very good coherence with the  transient period of 6820 sec is shown in  Fig. \ref{fig:vela6820modulo}.

\section{Discussion}   
\label{sec:discussion}

We have simulated the accretion flow in Vela X-1 with high spatial and temporal resolutions around the neutron star and found flares and off-states qualitatively in agreement with the observed ones. In particular off-states are regularly produced corresponding to an instability of the bow shock surrounding the neutron star.

In wind-fed HMXB systems \citep{1988ApJ...331L.117T} the accretion flows are complex and characterized by episodic accretion \citep{1990ApJ...358..545S,1991ApJ...376..750S} with an accreted angular momentum varying in direction and close to zero, on average. This is known as the `flip-flop' instability \citep[e.g.][]{1987MNRAS.226..785M,1988ApJ...327L..73T}, interpreted by the formation of transient accretion discs \citep{2009ApJ...700...95B}. We also observed the accreted angular momentum to change sign regularly in our simulations between flares and off-states but cannot characterize them with the usual `flip-flop' instability as the characteristic duration of the variability is much shorter in our case and not linked to the formation of accretion discs.

Our simulations predicts the formation of low density bubbles behind the bow shock, around the neutron star, resulting in the X-ray off-states. These bubbles are $\sim$ 10 times larger than the  Bondi-Hoyle-Lyttleton \citep[BHL; see e.g.,][]{2004NewAR..48..843E}  accretion radius ($\sim 10^{10}$ cm). The bow shock is not stable, it appears close to the neutron star and moves away up to a distance ($\ga 10^{11}$ cm; see right panel of  Fig. \ref{fig:map_offstate}) before gradually 
falling back when a stream of gas eventually reaches the neutron star producing a new rise of the X-ray flux. The accretion stream can either move left-handed or right-handed. 

This `breathing' behaviour is not perfectly periodic nor continuous but can be observed most of the time in the simulation and is at the origin of the off-states.
As the internal energy (pressure) of the low density bubble is a fraction  ($\sim$ 1/10) of the gravitational potential at the bow shock, 
the modulation period  ($\sim 6500 - 7000$ sec) is comparable to the free-fall time ($\sim$ 2000 sec) at this position.  
Neither the free-fall time from the  BHL accretion radius ($t_{ff}\sim 2$ minute) nor from the magnetospheric radius  ($t_{ff}\sim 3$ sec) are consistent with the observed 
 modulation time-scale.

Modulations are also produced by recent idealized 3D BHL accretion simulations \citep{2012ApJ...752...30B}, although weaker than in the 2D case \citep{1994ApJ...427..342R,1997A&A...317..793R,1999A&A...346..861R}. 
The time-scale of these  modulations is related to the accretion radius and their amplitude is weaker than what we are observing.

\citet{2006ApJ...638..369K} showed that the Bondi-Hoyle accretion of super-sonic turbulent gas has significant density enhancement when compared to the simple BHL formulation. They obtained a log-normal distribution of mass accretion rate, however much wider than what we found. The source of the turbulence is completely different in their case. The simulated log normal distribution features more low luminosity excess than observed. \citet{1998PhRvE..58.4501P} found similar excesses in one dimensional simulations of highly compressible gas. It is unclear if the additional excesses have a physical meaning. 

Reality is certainly more complex, indeed massive stars are known to have clumpy winds \citep{Owocki+88,2003A&A...406L...1D,2006MNRAS.372..313O,2013MNRAS.428.1837S} and a rich phenomenology might be produced \citep{Walter07Winds}. Clumpy wind scenarios can enhance the instabilities and reduce the X-ray luminosity for extended period of time, in the class of Super-giant Fast X-ray Transient \citep[SFXT,][]{2006ESASP.604..165N}. One dimensional studies of the influence of strong density and velocity fluctuations of the wind resulted in episodic X-ray variability, however too large to describe the observations \citep{2012MNRAS.421.2820O}. The lack of  multi-scale and multi-dimensional hydrodynamic simulations of macroscopic clumpy wind in a binary system does not yet allow to understand the interplay between intrinsic clumping and the various effects of the neutron star.

\section{Conclusion}   
\label{sec:conclude}
 
We compared hard X-ray light-curves of Vela X-1 obtained with RXTE and INTEGRAL with the predictions of hydrodynamic simulations. 

The simulated light-curve is highly variable (see  Fig. \ref{fig:velaLC}) as observed \citep{Kreykenbohm+08}. The dynamical range of variability is of the order of $\sim 10^{3}$ 
between off-states and the brightest flares. The X-ray luminosity is a direct probe of the instantaneous mass-accretion rate and of the density fluctuations around the neutron star. The duration of the off-states in our simulation (1 to 5 ksec) corresponds to the free fall time of low density bubbles building up behind the bow shock surrounding the neutron star.
 
The log normal flux distribution, resulting from the observations and from the simulation, is a characteristic result of self organised criticality \citep{1988PhRvA..38..364B,Crow1988,2005MNRAS.359..345U}. In our case the criticality condition is probably related to the direction of the bow shock and accretion stream that can lead or trail the neutron star.  Oscillations between these positions lead to the succession of off-states, flares, and more generally to the  near- log-normal flux distribution. The flares correspond to the accretion of a mass of $\sim 10^{19}$ gr, much smaller than inferred for the clumps in the case of supergiant fast X-ray transients \citep[SFTXs,][]{2011A&A...531A.130B,Walter07Winds}, because the variability is driven by small scale instabilities in Vela X-1.

Our hydrodynamic simulations are sufficient to explain the observed behaviour without the need for intrinsically clumpy stellar wind or high magnetic fields and gating mechanisms. More advanced and realistic simulations, including such phenomena are needed to understand their interplay with the hydrodynamic effects of the neutron star and to reveal the full  accretion phenomenology in classical sgHMXBs.


\begin{acknowledgements}
AM would like to thank Prof. J. Blondin for fruitful 
discussions and hospitality at the NCSU, as well as acknowledge support 
 by the Polish NCN grant  2012/04/M/ST9/00780.

\end{acknowledgements}

\bibliographystyle{aa} 
\bibliography{references} 

\begin{thebibliography}{54}
\expandafter\ifx\csname natexlab\endcsname\relax\def\natexlab#1{#1}\fi

\bibitem[{{Abbott}(1982)}]{1982ApJ...259..282A}
{Abbott}, D.~C. 1982, \apj, 259, 282

\bibitem[{{Bak} {et~al.}(1988){Bak}, {Tang}, \&
  {Wiesenfeld}}]{1988PhRvA..38..364B}
{Bak}, P., {Tang}, C., \& {Wiesenfeld}, K. 1988, \pra, 38, 364

\bibitem[{{Bildsten} {et~al.}(1997){Bildsten}, {Chakrabarty}, {Chiu}, {Finger},
  {Koh}, {Nelson}, {Prince}, {Rubin}, {Scott}, {Stollberg}, {Vaughan},
  {Wilson}, \& {Wilson}}]{1997ApJS..113..367B}
{Bildsten}, L., {Chakrabarty}, D., {Chiu}, J., {et~al.} 1997, \apjs, 113, 367

\bibitem[{{Blondin} {et~al.}(1990){Blondin}, {Kallman}, {Fryxell}, \&
  {Taam}}]{Blondin90}
{Blondin}, J.~M., {Kallman}, T.~R., {Fryxell}, B.~A., \& {Taam}, R.~E. 1990,
  \apj, 356, 591

\bibitem[{{Blondin} \& {Pope}(2009)}]{2009ApJ...700...95B}
{Blondin}, J.~M. \& {Pope}, T.~C. 2009, \apj, 700, 95

\bibitem[{{Blondin} \& {Raymer}(2012)}]{2012ApJ...752...30B}
{Blondin}, J.~M. \& {Raymer}, E. 2012, \apj, 752, 30

\bibitem[{{Blondin} {et~al.}(1991){Blondin}, {Stevens}, \&
  {Kallman}}]{Blondin91}
{Blondin}, J.~M., {Stevens}, I.~R., \& {Kallman}, T.~R. 1991, \apj, 371, 684

\bibitem[{{Bozzo} {et~al.}(2011){Bozzo}, {Giunta}, {Cusumano}, {Ferrigno},
  {Walter}, {Campana}, {Falanga}, {Israel}, \& {Stella}}]{2011A&A...531A.130B}
{Bozzo}, E., {Giunta}, A., {Cusumano}, G., {et~al.} 2011, \aap, 531, A130+

\bibitem[{{Castor} {et~al.}(1975){Castor}, {Abbott}, \& {Klein}}]{CAKwind}
{Castor}, J.~I., {Abbott}, D.~C., \& {Klein}, R.~I. 1975, \apj, 195, 157

\bibitem[{{Colella} \& {Woodward}(1984)}]{PPMCW}
{Colella}, P. \& {Woodward}, P.~R. 1984, Journal of Computational Physics, 54,
  174

\bibitem[{{Crow} \& {Shimizu}(1988)}]{Crow1988}
{Crow}, E.~L. \& {Shimizu}, K. 1988, {Lognormal Distributions: Theory and
  Applictions} ({Dekker, New York})

\bibitem[{{Dessart} \& {Owocki}(2003)}]{2003A&A...406L...1D}
{Dessart}, L. \& {Owocki}, S.~P. 2003, \aap, 406, L1

\bibitem[{{Doroshenko} {et~al.}(2011){Doroshenko}, {Santangelo}, \&
  {Suleimanov}}]{2011A&A...529A..52D}
{Doroshenko}, V., {Santangelo}, A., \& {Suleimanov}, V. 2011, \aap, 529, A52

\bibitem[{{Dupree} {et~al.}(1980){Dupree}, {Gursky}, {Black}, {Davis},
  {Hartmann}, {Matilsky}, {Raymond}, {Hammerschlag-Hensberge}, {van den
  Heuvel}, {Burger}, {Lamers}, {Vanden Bout}, {Morton}, {De Loore}, {van
  Dessel}, {Menzies}, {Whitelock}, {Watson}, {Sanford}, \&
  {Pollard}}]{1980ApJ...238..969D}
{Dupree}, A.~K., {Gursky}, H., {Black}, J.~H., {et~al.} 1980, \apj, 238, 969

\bibitem[{{Edgar}(2004)}]{2004NewAR..48..843E}
{Edgar}, R. 2004, \nar, 48, 843

\bibitem[{{Fransson} \& {Fabian}(1980)}]{FarnssonFabian1980}
{Fransson}, C. \& {Fabian}, A.~C. 1980, \aap, 87, 102

\bibitem[{{Friend} \& {Abbott}(1986)}]{1986ApJ...311..701F}
{Friend}, D.~B. \& {Abbott}, D.~C. 1986, \apj, 311, 701

\bibitem[{{F{\"u}rst} {et~al.}(2010){F{\"u}rst}, {Kreykenbohm}, {Pottschmidt},
  {Wilms}, {Hanke}, {Rothschild}, {Kretschmar}, {Schulz}, {Huenemoerder},
  {Klochkov}, \& {Staubert}}]{Furstetal10}
{F{\"u}rst}, F., {Kreykenbohm}, I., {Pottschmidt}, K., {et~al.} 2010, ArXiv
  e-prints

\bibitem[{{Hunt}(1971)}]{1971MNRAS.154..141H}
{Hunt}, R. 1971, \mnras, 154, 141

\bibitem[{{Illarionov} \& {Sunyaev}(1975)}]{1975A&A....39..185I}
{Illarionov}, A.~F. \& {Sunyaev}, R.~A. 1975, \aap, 39, 185

\bibitem[{{Jahoda} {et~al.}(2006){Jahoda}, {Markwardt}, {Radeva}, {Rots},
  {Stark}, {Swank}, {Strohmayer}, \& {Zhang}}]{PCA-ref2}
{Jahoda}, K., {Markwardt}, C.~B., {Radeva}, Y., {et~al.} 2006, \apjs, 163, 401

\bibitem[{{Kallman} \& {McCray}(1982)}]{Kallman82}
{Kallman}, T.~R. \& {McCray}, R. 1982, \apjs, 50, 263

\bibitem[{{Kreykenbohm} {et~al.}(1999){Kreykenbohm}, {Kretschmar}, {Wilms},
  {Staubert}, {Kendziorra}, {Gruber}, {Heindl}, \&
  {Rothschild}}]{1999A&A...341..141K}
{Kreykenbohm}, I., {Kretschmar}, P., {Wilms}, J., {et~al.} 1999, \aap, 341, 141

\bibitem[{{Kreykenbohm} {et~al.}(2008){Kreykenbohm}, {Wilms}, {Kretschmar},
  {Torrej{\'o}n}, {Pottschmidt}, {Hanke}, {Santangelo}, {Ferrigno}, \&
  {Staubert}}]{Kreykenbohm+08}
{Kreykenbohm}, I., {Wilms}, J., {Kretschmar}, P., {et~al.} 2008, \aap, 492, 511

\bibitem[{{Krti{\v c}ka} {et~al.}(2012){Krti{\v c}ka}, {Kub{\'a}t}, \&
  {Skalick{\'y}}}]{2012ApJ...757..162K}
{Krti{\v c}ka}, J., {Kub{\'a}t}, J., \& {Skalick{\'y}}, J. 2012, \apj, 757, 162

\bibitem[{Krumholz {et~al.}(2006)Krumholz, McKee, \&
  Klein}]{2006ApJ...638..369K}
Krumholz, M.~R., McKee, C.~F., \& Klein, R.~I. 2006, The Astrophysical Journal,
  638, 369

\bibitem[{{Kudritzki} \& {Puls}(2000)}]{winds_from_hot_stars}
{Kudritzki}, R. \& {Puls}, J. 2000, \araa, 38, 613

\bibitem[{{Lebrun} {et~al.}(2003){Lebrun}, {Leray}, {Lavocat}, {Cr{\'e}tolle},
  {Arqu{\`e}s}, {Blondel}, {Bonnin}, {Bou{\`e}re}, {Cara}, {Chaleil}, {Daly},
  {Desages}, {Dzitko}, {Horeau}, {Laurent}, {Limousin}, {Mathy}, {Mauguen},
  {Meignier}, {Molini{\'e}}, {Poindron}, {Rouger}, {Sauvageon}, \&
  {Tourrette}}]{isgri-ref}
{Lebrun}, F., {Leray}, J.~P., {Lavocat}, P., {et~al.} 2003, \aap, 411, L141

\bibitem[{{Matsuda} {et~al.}(1987){Matsuda}, {Inoue}, \&
  {Sawada}}]{1987MNRAS.226..785M}
{Matsuda}, T., {Inoue}, M., \& {Sawada}, K. 1987, \mnras, 226, 785

\bibitem[{{Nagase}(1989)}]{1989PASJ...41....1N}
{Nagase}, F. 1989, \pasj, 41, 1

\bibitem[{{Nagase} {et~al.}(1986){Nagase}, {Hayakawa}, {Sato}, {Masai}, \&
  {Inoue}}]{1986PASJ...38..547N}
{Nagase}, F., {Hayakawa}, S., {Sato}, N., {Masai}, K., \& {Inoue}, H. 1986,
  \pasj, 38, 547

\bibitem[{{Negueruela} {et~al.}(2006){Negueruela}, {Smith}, {Reig}, {Chaty}, \&
  {Torrej{\'o}n}}]{2006ESASP.604..165N}
{Negueruela}, I., {Smith}, D.~M., {Reig}, P., {Chaty}, S., \& {Torrej{\'o}n},
  J.~M. 2006, in ESA Special Publication, Vol. 604, The X-ray Universe 2005,
  ed. {A.~Wilson}, 165--+

\bibitem[{{Oskinova} {et~al.}(2006){Oskinova}, {Feldmeier}, \&
  {Hamann}}]{2006MNRAS.372..313O}
{Oskinova}, L.~M., {Feldmeier}, A., \& {Hamann}, W.-R. 2006, \mnras, 372, 313

\bibitem[{{Oskinova} {et~al.}(2012){Oskinova}, {Feldmeier}, \&
  {Kretschmar}}]{2012MNRAS.421.2820O}
{Oskinova}, L.~M., {Feldmeier}, A., \& {Kretschmar}, P. 2012, \mnras, 421, 2820

\bibitem[{{Owocki} {et~al.}(1988){Owocki}, {Castor}, \& {Rybicki}}]{Owocki+88}
{Owocki}, S.~P., {Castor}, J.~I., \& {Rybicki}, G.~B. 1988, \apj, 335, 914

\bibitem[{{Owocki} \& {Rybicki}(1984)}]{1984ApJ...284..337O}
{Owocki}, S.~P. \& {Rybicki}, G.~B. 1984, \apj, 284, 337

\bibitem[{{Passot} \& {V{\'a}zquez-Semadeni}(1998)}]{1998PhRvE..58.4501P}
{Passot}, T. \& {V{\'a}zquez-Semadeni}, E. 1998, \pre, 58, 4501

\bibitem[{{Quaintrell} {et~al.}(2003){Quaintrell}, {Norton}, {Ash}, {Roche},
  {Willems}, {Bedding}, {Baldry}, \& {Fender}}]{Quaintrell_et_al03}
{Quaintrell}, H., {Norton}, A.~J., {Ash}, T.~D.~C., {et~al.} 2003, \aap, 401,
  313

\bibitem[{Ruffert(1994)}]{1994ApJ...427..342R}
Ruffert, M. 1994, Astrophysical Journal, 427, 342

\bibitem[{Ruffert(1997)}]{1997A&A...317..793R}
Ruffert, M. 1997, Astronomy and Astrophysics, 317, 793

\bibitem[{Ruffert(1999)}]{1999A&A...346..861R}
Ruffert, M. 1999, Astronomy and Astrophysics, 346, 861

\bibitem[{{Shakura} {et~al.}(2013){Shakura}, {Postnov}, \&
  {Hjalmarsdotter}}]{2013MNRAS.428..670S}
{Shakura}, N., {Postnov}, K., \& {Hjalmarsdotter}, L. 2013, \mnras, 428, 670

\bibitem[{{Shakura} {et~al.}(2012){Shakura}, {Postnov}, {Kochetkova}, \&
  {Hjalmarsdotter}}]{2012MNRAS.420..216S}
{Shakura}, N., {Postnov}, K., {Kochetkova}, A., \& {Hjalmarsdotter}, L. 2012,
  \mnras, 420, 216

\bibitem[{{Soker}(1990)}]{1990ApJ...358..545S}
{Soker}, N. 1990, \apj, 358, 545

\bibitem[{{Soker}(1991)}]{1991ApJ...376..750S}
{Soker}, N. 1991, \apj, 376, 750

\bibitem[{{Stevens} \& {Kallman}(1990)}]{1990ApJ...365..321S}
{Stevens}, I.~R. \& {Kallman}, T.~R. 1990, \apj, 365, 321

\bibitem[{{Sundqvist} \& {Owocki}(2013)}]{2013MNRAS.428.1837S}
{Sundqvist}, J.~O. \& {Owocki}, S.~P. 2013, \mnras, 428, 1837

\bibitem[{Taam {et~al.}(1988)Taam, Brown, \& Fryxell}]{1988ApJ...331L.117T}
Taam, R.~E., Brown, D.~A., \& Fryxell, B.~A. 1988, Astrophysical Journal, 331,
  L117

\bibitem[{Taam \& Fryxell(1988)}]{1988ApJ...327L..73T}
Taam, R.~E. \& Fryxell, B.~A. 1988, Astrophysical Journal, 327, L73

\bibitem[{{Tarter} {et~al.}(1969){Tarter}, {Tucker}, \&
  {Salpeter}}]{1969ApJ...156..943T}
{Tarter}, C.~B., {Tucker}, W.~H., \& {Salpeter}, E.~E. 1969, \apj, 156, 943

\bibitem[{{Uttley} {et~al.}(2005){Uttley}, {McHardy}, \&
  {Vaughan}}]{2005MNRAS.359..345U}
{Uttley}, P., {McHardy}, I.~M., \& {Vaughan}, S. 2005, \mnras, 359, 345

\bibitem[{{Walter} {et~al.}(2010){Walter}, {Rohlfs}, {Meharga}, {Binko},
  {Morisset}, {Beck}, {Produit}, {Pavan}, {Savchenko}, {Ferrigno},
  {Frankowski}, \& {Bordas}}]{heavens-ref}
{Walter}, R., {Rohlfs}, R., {Meharga}, M.~T., {et~al.} 2010, in Proceedings of
  the 8th INTEGRAL Workshop ''The Restless Gamma-ray Universe'' (INTEGRAL
  2010), 162.

\bibitem[{{Walter} \& {Zurita Heras}(2007)}]{Walter07Winds}
{Walter}, R. \& {Zurita Heras}, J. 2007, \aap, 476, 335

\bibitem[{{Winkler} {et~al.}(2003){Winkler}, {Courvoisier}, {Di Cocco},
  {Gehrels}, {Gim{\'e}nez}, {Grebenev}, {Hermsen}, {Mas-Hesse}, {Lebrun},
  {Lund}, {Palumbo}, {Paul}, {Roques}, {Schnopper}, {Sch{\"o}nfelder},
  {Sunyaev}, {Teegarden}, {Ubertini}, {Vedrenne}, \& {Dean}}]{integral-ref}
{Winkler}, C., {Courvoisier}, T., {Di Cocco}, G., {et~al.} 2003, \aap, 411, L1

\end{thebibliography}

\end{document}